\documentclass[conference]{IEEEtran}

\usepackage{algorithmic}
\usepackage{amsmath,amssymb,amsfonts}
\usepackage{balance}
\usepackage{bm}
\usepackage{booktabs}
\usepackage{cite}
\usepackage{colortbl}
\usepackage{float}
\usepackage[T1]{fontenc}
\usepackage{graphicx}
\usepackage{makecell}
\usepackage{mathtools}
\usepackage{multirow}
\usepackage{siunitx}
\usepackage{soul}
\usepackage{subfig}
\usepackage{textcomp}
\usepackage{threeparttable}
\usepackage[color=blue!40]{todonotes}
\usepackage{xcolor}
\usepackage{hyperref}
\hypersetup{bookmarksnumbered, colorlinks=false, pdfborder={0 0 1}}%
\soulregister\footnote7
\soulregister\eqref7
\soulregister\cite7
\soulregister\ref7
\graphicspath{{./}{../fig/}}

\IEEEoverridecommandlockouts

\begin{document}

\title{%
  Impact of Environmental Factors on LoRa \SI{2.4}{\GHz} Time of Flight Ranging
  Outdoors}%
\author{%
   \IEEEauthorblockN{%
     Yiqing Zhou$^{1}$, Xule Zhou$^{1}$, Zecan Cheng$^{1}$, Chenao Lu$^{1}$,
     Junhan Chen$^{1}$, Jiahong Pan$^{1}$, Yizhuo Liu$^{1}$,\\
     Sihao Li$^{2,\star}$, and Kyeong Soo Kim$^{1,\star}$}%
   \IEEEauthorblockA{%
     $^{1}$~School of Advanced Technology, Xi'an Jiaotong-Liverpool University,
     Suzhou, 215123, P.R. China.\\
     Email: \{Yiqing.Zhou24, Xule.Zhou23, Zecan.Cheng24, Chenao.Lu23,
     Junhan.Chen23,\\
     Jiahong.Pan22, Yizhou.Liu23\}@student.xjtlu.edu.cn, Kyeongsoo.Kim@xjtlu.edu.cn\\
     $^{2}$~School of Artificial Intelligence, Suzhou Vocational
     Institute of Industrial Technology, Suzhou 215104, P.R. China.\\
     Email: 01177@siit.edu.cn}%
   \thanks{$\star$~Corresponding authors.}%
}%

\maketitle

\begin{abstract}
  In WSN/IoT, node localization is essential to long-running applications for
  accurate environment monitoring and event detection, often covering a large
  area in the field. Due to the lower time resolution of typical WSN/IoT
  platforms (e.g., \SI{1}{\us} on ESP32 platforms) and the jitters in
  timestamping, packet-level localization techniques cannot provide meter-level
  resolution. For high-precision localization as well as world-wide
  interoperability via 2.4-\si{\GHz} ISM band, a new variant of LoRa, called
  LoRa \SI{2.4}{\GHz}, was proposed by semtech, which provides a radio frequency
  (RF) time of flight (ToF) ranging method for meter-level
  localization. However, the existing datasets reported in the literature are
  limited in their coverages and do not take into account varying environmental
  factors such as temperature and humidity. To address these issues, LoRa
  \SI{2.4}{\GHz} RF ToF ranging data was collected on a sports field at the
  XJTLU south campus, where three LoRa nodes logged samples of ranging with a
  LoRa base station, together with temperature and humidity, at reference points
  arranged as a $3{\times}3$ grid covering \SI{400}{\m\squared} over three weeks
  and uploaded all measurement records to the base station equipped with an
  ESP32-based transceiver for machine and user communications. The results of a
  preliminary investigation based on a simple deep neural network (DNN) model
  demonstrate that the environmental factors, including the temperature and
  humidity, significantly affect the accuracy of ranging, which calls for
  advanced methods of compensating for the effects of environmental factors on
  LoRa RF ToF ranging outdoors.
\end{abstract}


\section{Introduction}
\label{sec:intro}
LoRa is one of the most popular technologies for low-power wide-area network
(LPWAN), which can transmit data packets over several kilometers at low data
rates with low-power consumption for a large variety of wireless sensor network
(WSN)/Internet of Things (IoT) applications. Conventional LoRa implementations
use a spread spectrum technology on license-free sub-\si{\GHz} bands, which are
different from one geographical region to another, resulting in many different
parameter settings including frequency channel, maximum duty-cycle, and maximum
transmission power. LoRa \SI{2.4}{\GHz} recently proposed by
Semtech~\cite{semtech:sx1280_sx1281}, on the other hand, uses the common
2.4-\si{\GHz} ISM band and thereby enables a worldwide coverage using a single
platform without changing system parameters~\cite{derevianckine23:_oppor_lora}.

Of the several features of LoRa \SI{2.4}{\GHz}, the high-precision ranging
implemented in Semtech's SX1280 chip can provide meter-level localization (i.e.,
up to \SI{2}{\m}) thanks to its use of radio frequency (RF) time of flight
(ToF)~\cite{semtech:ranging,lanzisera06:_rf}, which is not possible in
conventional packet-level localization techniques due to the lower time
resolution of typical WSN/IoT platforms (e.g., \SI{1}{\us} on ESP32 platforms)
and the jitters in timestamping. The introduction of the high-precision RF ToF
ranging could have a significant impact on the localization of WSN/IoT nodes
deployed in wide areas; unlike the existing approaches for global positioning
systems (GPS)-free geolocation based on LoRa, which suffer from low localization
accuracy resulting from either conventional packet-level ToF or received signal
strength (RSS)/received signal strength indicator (RSSI) fingerprinting, the
GPS-free geolocation based on RF ToF could provide localization accuracy similar
to or even better than that of GPS. Though the performance of Lora
\SI{2.4}{\GHz} RF ToF ranging has been evaluated in terms of SX1280 transceiver
parameter settings~\cite{rander20:_rangin_lora}, the impact of environmental
factors, like temperature and humidity, and their compensation techniques are
yet to be fully investigated, which is important because LoRa nodes are
typically deployed in outdoor environments.

In this paper, we present the results of our preliminary investigation of the
impact of the environmental factors of temperature and humidity on the
performance of LoRa \SI{2.4}{\GHz} RF ToF ranging outdoors, which is based on
the ranging data collected together with temperature and humidity on a sports
field at the XJTLU south campus over three weeks. We also explore the potential
of compensating for the impact of environmental factors on LoRa RF ToF ranging
based on a simple deep neural network (DNN) model

The rest of the paper is organized as follows: Section~\ref{sec:related-work}
reviews related work. Section~\ref{sec:methodology} describes our methodology
for the investigation of LoRa \SI{2.4}{\GHz} RF ToF
ranging. Section~\ref{sec:experimental-results} presents the results of our
preliminary investigation based on the constructed database and a simple DNN
model. Section~\ref{sec:conclusions} concludes our work.

\section{Related Work}
\label{sec:related-work}

\subsection{GPS-Free Localization Techniques for WSN/IoT}
\label{sec:gps-free-localization}
WSN/IoT localization techniques not based on GPS can be grouped into those with
and without ranging~\cite{he16:_wi_fi}. Ranging-based approaches rely on the
distances between a node location and multiple anchor nodes with known locations
or differences of them in determining the unknown node location through
trilateration/multilateration. The ToF---also called time of arrival
(ToA)---technique uses the travel time between a node location and anchor
nodes~\cite{golden07:_sensor_wi_fi,chen16:_toa_imu_kalman_map}, while the time
difference of arrival (TDoA) technique uses the time differences between the
arrivals of the signals from a node at anchor
nodes~\cite{golden07:_sensor_wi_fi,chen16:_toa_imu_kalman_map}. Instead of
arrival times or their differences, the angle of arrival (AoA) technique uses
the angles of signal arrivals, which can be estimated by measuring time
differences of arrivals between individual elements of an antenna
array~\cite{wen14:_aoa_ieee,xiong13:_array}.

In ranging-free localization techniques, a node location is not estimated based
on distance-related information with trilateration/multilateration. In the case
of \textit{location fingerprinting}, the information measured at a node, e.g.,
RSSIs, channel state information (CSI), and geomagnetic field intensity, is used
for localization, which is supposed to be unique to each location and thereby
serves as a location fingerprint. RSSI fingerprinting, for example, has two
operation phases of offline and
online~\cite{xiao16:_survey_wirel_indoor_local_devic_persp}: During the offline
phase, the RSSIs at known locations, called reference points (RPs), are
collected and stored in a fingerprint database; during the online phase, the
node location is estimated based on the RPs whose RSSIs most closely match the
newly-measured RSSIs at the node. In the case of Wi-Fi networks, CSI also can be
used as location fingerprints, which, unlike coarse-grained RSSIs, can provide
fine-grained indicators consisting of both amplitude and phase information
during signal propagation and enables even single-access point (AP)
localization. A significant drawback of CSI-based Wi-Fi fingerprinting, however,
is the requirement of unique network interface cards (NICs) and device drivers
for the acquisition of CSI (e.g., Intel 5300 NICs)~\cite{zhang19:_csi_ap}.


\subsection{RF ToF Ranging}
\label{sec:rf-tof}
As mentioned in Section~\ref{sec:intro}, the performance of the conventional
packet-level ToF ranging is significantly limited by the lower time resolution
of typical WSN/IoT platforms. With \SI{1}{\us} clock resolution, ToF distance
estimate error can be up to \SI{2000}{\m} on LoPy4 development boards mounted on
PyTrack sensor shields~\cite{danebjer18:_gps_lora}.

RF ToF ranging can address the limitation of the conventional ToF ranging and
provide meter-level accuracy on comparable hardware platforms with little
overhead of simple signal processing blocks~\cite{thorbjornsen10:_radio_rf}. In
bandwidth limited systems like WSN/IoT, sub-clock ToF measurements can be done
by resolving the phase offset of a signal. For example, pseudorandom noise (PN)
codes can be used as signals for measuring small phase offsets because the
autocorrelation function of a PN code exhibits a single large peak moving with a
phase offset, which is useful because a sequence of $N$ values can be converted
into a single feature with an effective signal to noise ratio (SNR) $N$ times
larger than that of the values used to construct it. The SNR enhancement is
advantageous for noise limited ranging because interference from other signals,
noise, or multipath propagation is a primary source of errors in RF ToF
ranging~\cite{lanzisera06:_rf}.

\section{Methodology}
\label{sec:methodology}
\begin{figure}[!htb]
  \centering%
  \includegraphics[width=\linewidth]{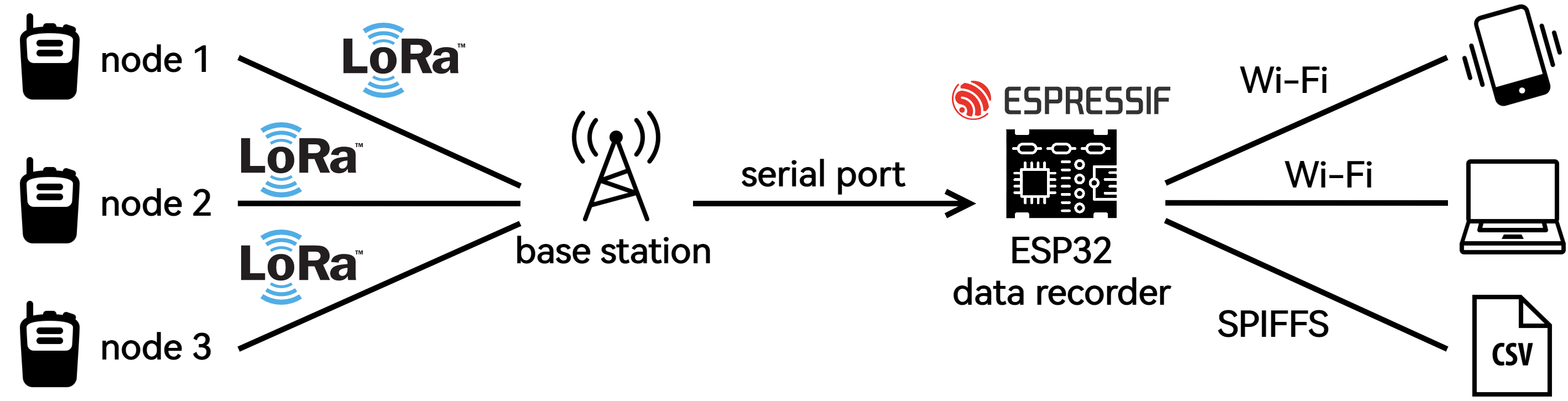}
  \caption{A network architecture for LoRa RF ToF ranging experiments.}
  \label{fig:network-architecture}
\end{figure}

A network architecture for LoRa RF ToF ranging experiments is shown in
\autoref{fig:network-architecture}, where an ESP32 board (i.e., WeMos D1 mini)
facilitates wireless communication between end-users and LoRa nodes through a
base station. While carrying out ranging, the LoRa nodes also capture
temperature and humidity data using AHT20 sensors and log the measured data with
corresponding timestamps and two-dimensional (2D) coordinates of RPs in a CSV
file. A built-in web interface of the ESP32 board allows end-users to view live
serial output and provides options to download or delete the CSV file. When
Wi-Fi is unavailable, the time is calculated using the \texttt{millis()}
function returning the number of milliseconds since the device was booted.

For distance measurements, the LoRa nodes use the SEMTECH SX1280-based
transceiver at 2.4-\si{\GHz} ISM-band, featuring an integrated RF ToF ranging
engine. The ranging process involves a two-way exchange of packets between a
LoRa node (i.e., a master) and a base station (i.e., a slave), measuring the
round-trip time to estimate the distance. The SX1280 supports various
configurations, including bandwidth, spreading factor, and coding rate, which
can be adjusted to optimize performance based on environmental conditions. For
our experiments, we apply the default settings for ranging applications.


\section{Experimental Results}
\label{sec:experimental-results}

\subsection{Experimental Setup}
\label{sec:experimental-setup}
\begin{figure}[!htb]
  \centering%
  \includegraphics[width=.5\linewidth]{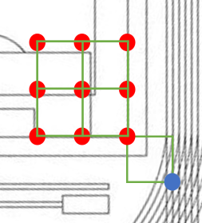}
  \caption{The arrangement of RPs (red circles) and a base station (a blue
    circle) on a 10-\si{\m} grid on a sports field at the XJTLU south campus.}
  \label{fig:reference-points}
\end{figure}

The RPs were established on a sports field at the XJTLU south campus
as shown in Fig.~\ref{fig:reference-points}. The nine red-circled RPs were
located at one of 10-\si{\m} grid points with the right lower RP as the origin
of the coordinates, i.e., from $(0,0)$ to $(20,20)$, while the blue-circled
base station was placed at $({-}10,{-}10)$.

Both a high-precision tape measure and a laser rangefinder were used for
marking the RPs on the field to ensure positional accuracy.

\begin{figure}[!htb]
  \centering%
  \includegraphics[width=.5\linewidth]{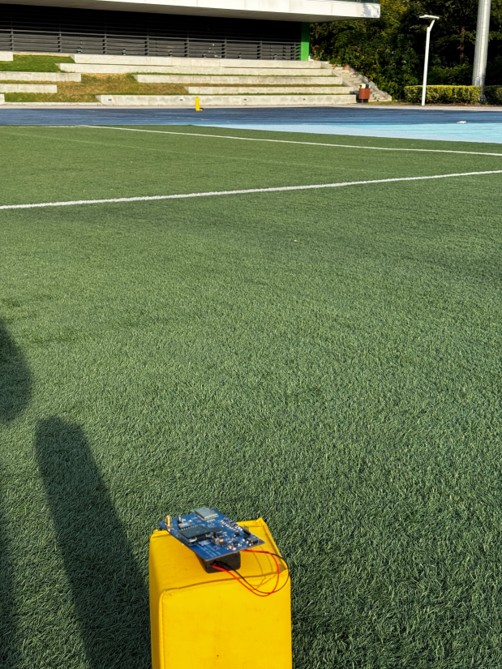}
  \caption{A LoRa node deployed in the field for multi-range mass data
    acquisition.}
  \label{fig:lora-node}
\end{figure}

For the experiments, four SX1280 LoRa evaluation boards were configured for RF
ToF ranging; one of them was designated as the base station, and the other
three were served as client nodes sequentially placed at each of the nine RPs
to perform ranging measurements. As shown in Fig.~\ref{fig:lora-node}, each
evaluation board was mounted atop a rigid post approximately \SI{40}{\cm} high.
This height was chosen to maintain consistency across measurements and to
reduce potential signal interference from the artificial turf surface.
Smartphones or laptops were used to download the collected data from the
evaluation boards. About 1,000 distinct ranging measurements were recorded at
each RP over a three-week observation period.


\subsection{Data Analysis}
\label{sec:data-analysis}
We first analyze the distribution of the LoRa RF ToF ranging errors over the
RPs.
For this purpose, we calculate the mean absolute ranging error at each RP after
outlier handling, which is defined as
\begin{equation}
  \bar{e}(x,y) = \frac{1}{N_{x,y}} \sum_{i \in \mathcal{G}(x,y)} e_i,
\end{equation}
where $\mathcal{G}(x,y)$ denotes a set of records collected at an RP located at
$(x,y)$, $N_{x,y}$ is its cardinality, and $e_{i}$ is a per-record ranging
error calculated as
\begin{equation}
  \label{eq:per-record-error}
  e_i = \bigl|\text{Measured Distance}_i - \text{Ground-Truth Distance}_i\bigr|.
\end{equation}
Note that the ground-truth distance in \eqref{eq:per-record-error} is
calculated based on the 2D coordinates of the RP and the base station.

\begin{figure}[!t]
  \centering
  \includegraphics[width=\linewidth,trim=40 40 30 60,clip]{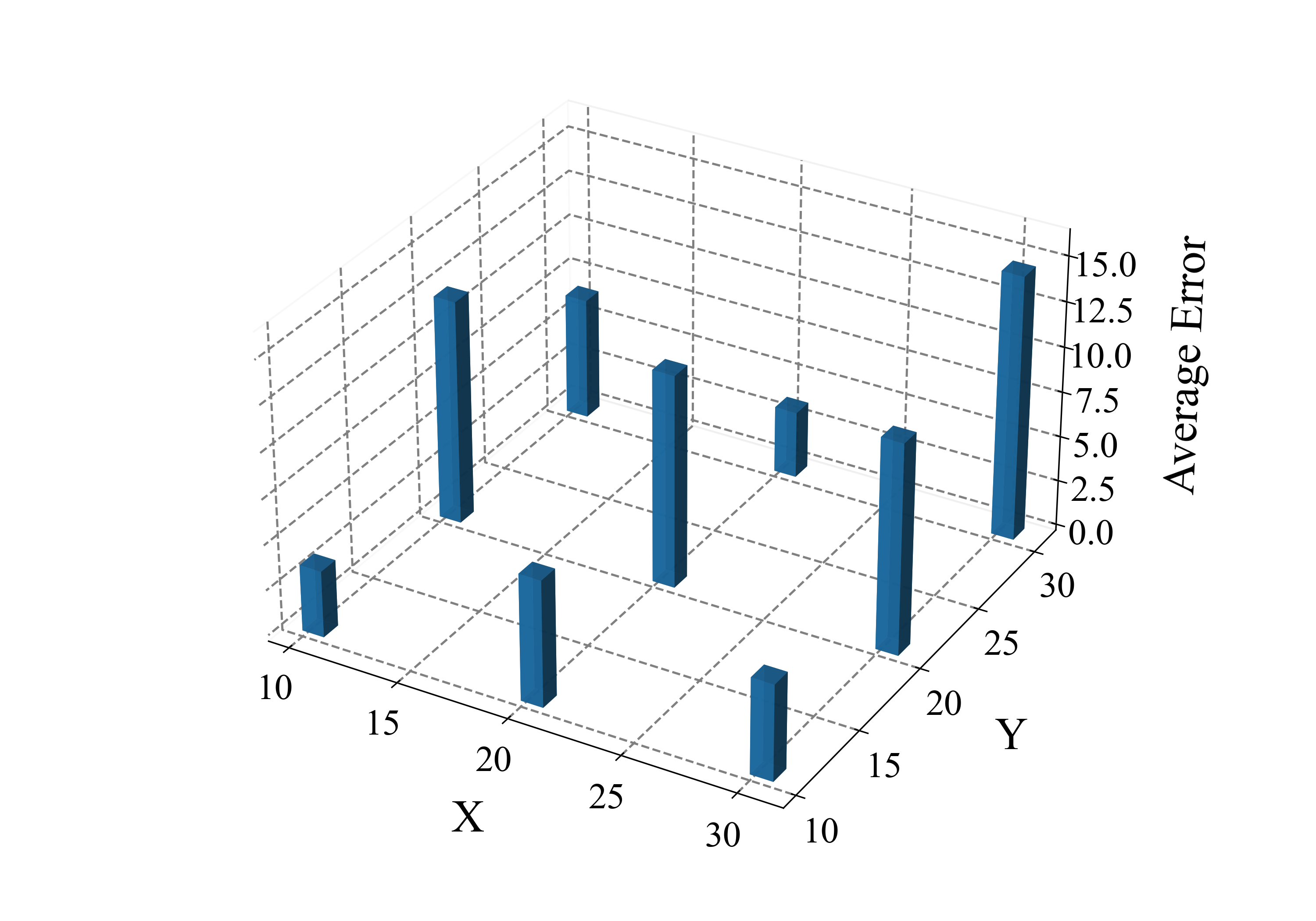}
  \caption{A 3D bar chart of the mean of the absolute ranging errors at each RP.}
  \label{fig:avg-error-3d}
\end{figure}

As shown in \autoref{fig:avg-error-3d}, the three-dimensional (3D) bar chart of
$\bar{e}(x,y)$ over all the RPs exhibit significant spatial variability of the
mean absolute ranging errors across RPs, which indicates that geometrical and
measurement conditions (e.g., relative orientation to the base station,
blockage, and multipath) can affect the ranging errors.

\begin{figure}[!t]
  \centering
  \includegraphics[width=\linewidth,trim=40 40 30 60,clip]{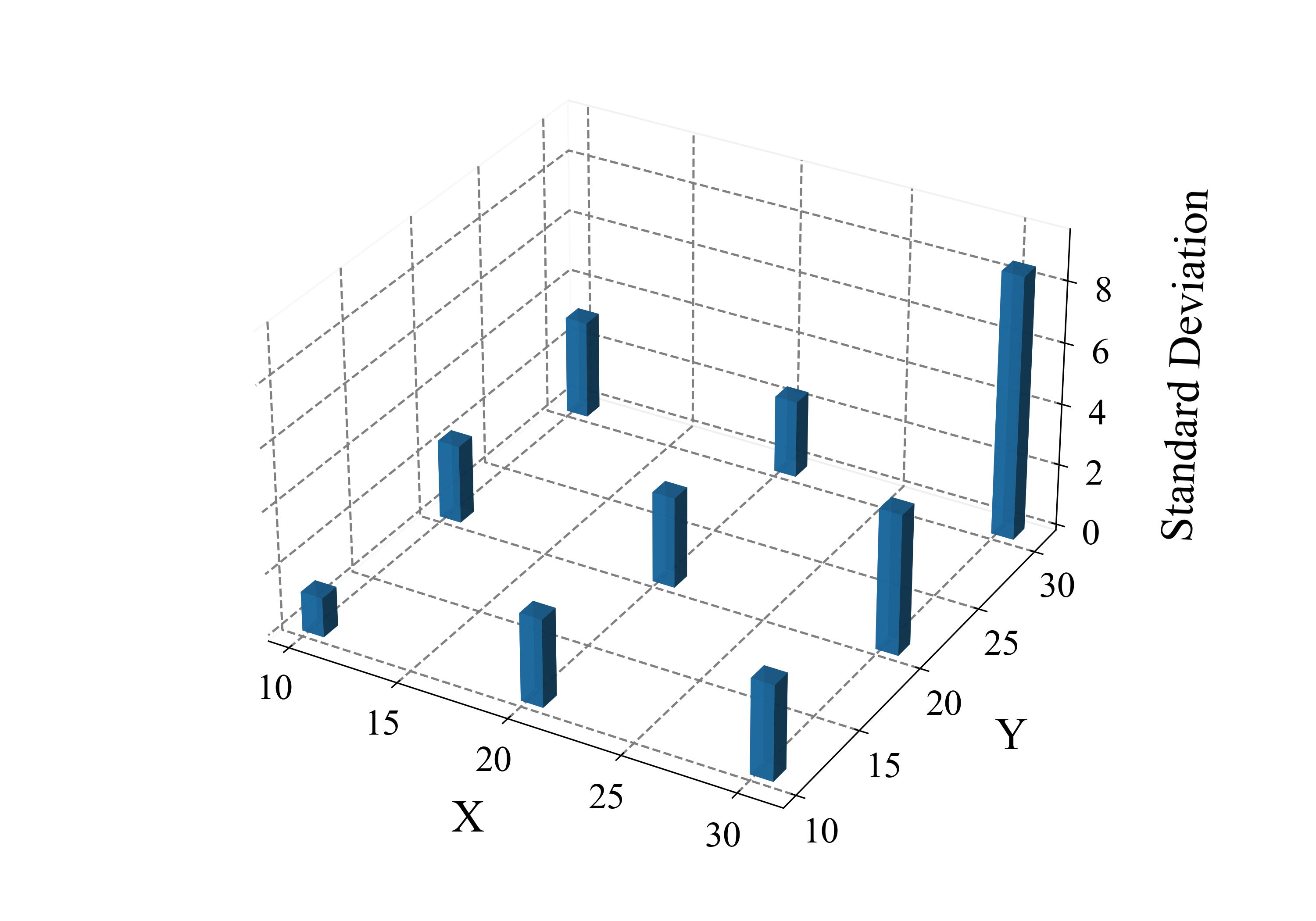}
  \caption{A 3D bar chart of the standard deviation of the absolute ranging
    errors at each RP.}
  \label{fig:std-error-3d}
\end{figure}

We also quantify the stability of ranging at each RP using the standard
deviation of the absolute ranging errors, which is shown in
\autoref{fig:std-error-3d}, where we can observe that the standard deviation at
the RP located at $(30,30)$ is significantly greater than the others,
indicating stronger fluctuations and thus poorer ranging stability at that RP;
the minimum standard deviation of \SI{1.27}{\m} is observed at the RP located
at $(10,10)$, while the majority of points fall in the 2.5--\SI{3.2}{\m} range.

\begin{figure}[!t]
  \centering
  \includegraphics[width=\linewidth]{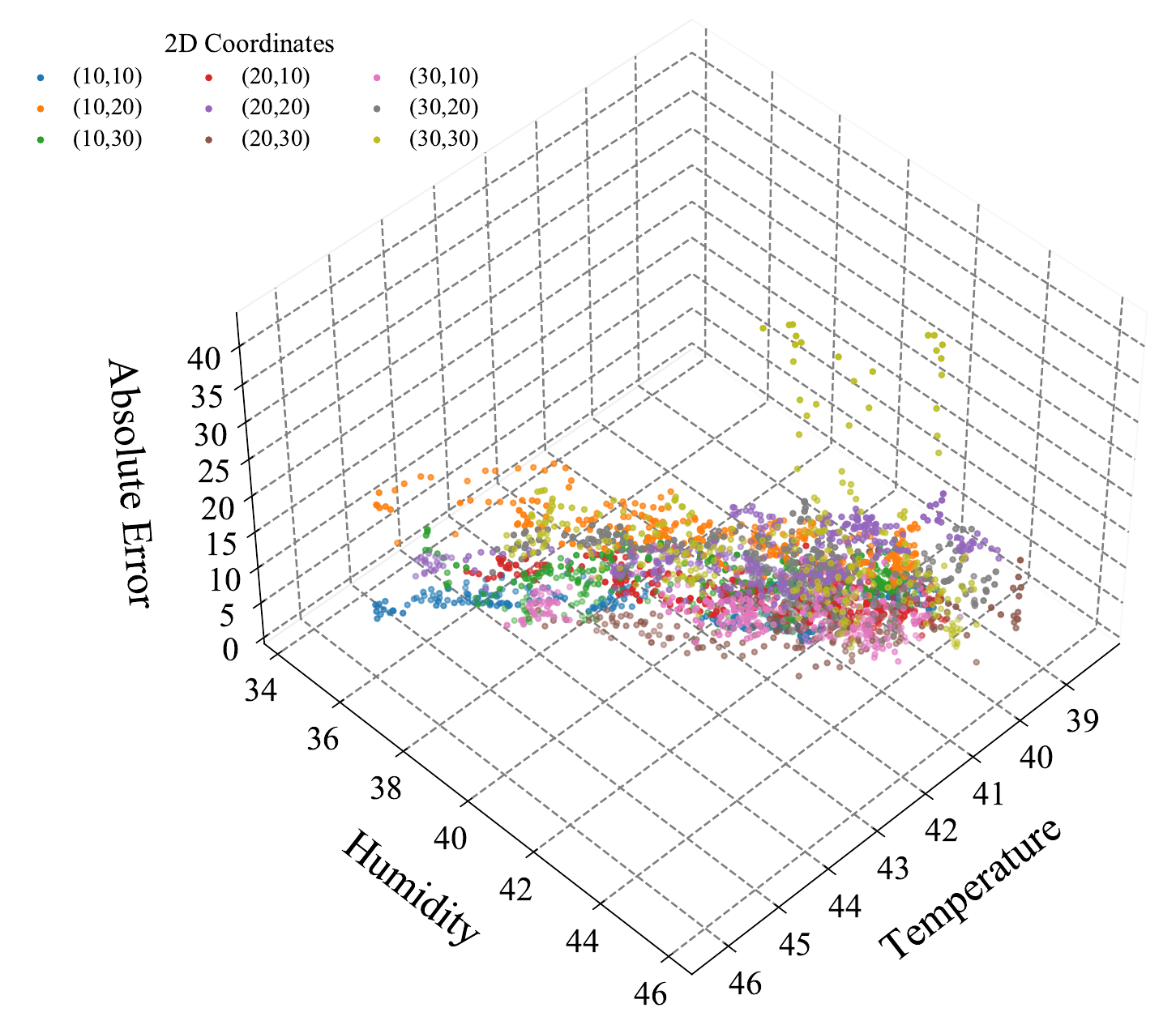}
  \caption{A 3D scatter plot of temperature ($X$)-humidity ($Y$)-absolute
    ranging error ($Z$).}
  \label{fig:3d-scatter-plot}
\end{figure}
\begin{figure}[!t]
  \centering
  \includegraphics[width=\linewidth,trim=0 100 0 120,clip]{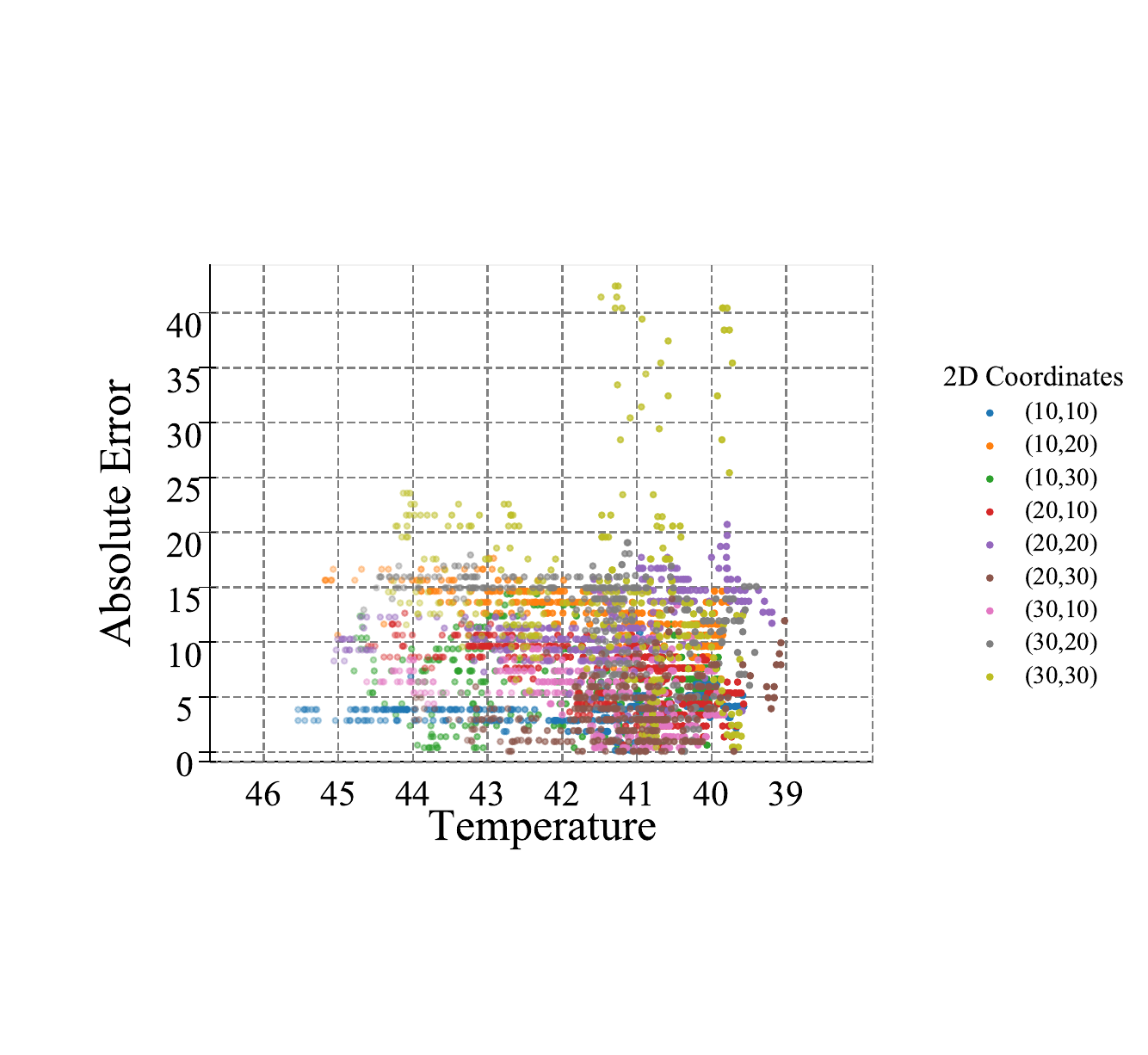}
  \caption{Absolute ranging error vs. temperature.}
  \label{fig:error-vs-temperature}
\end{figure}
\begin{figure}[!t]
  \centering
  \includegraphics[width=\linewidth,trim=0 100 0 120,clip]{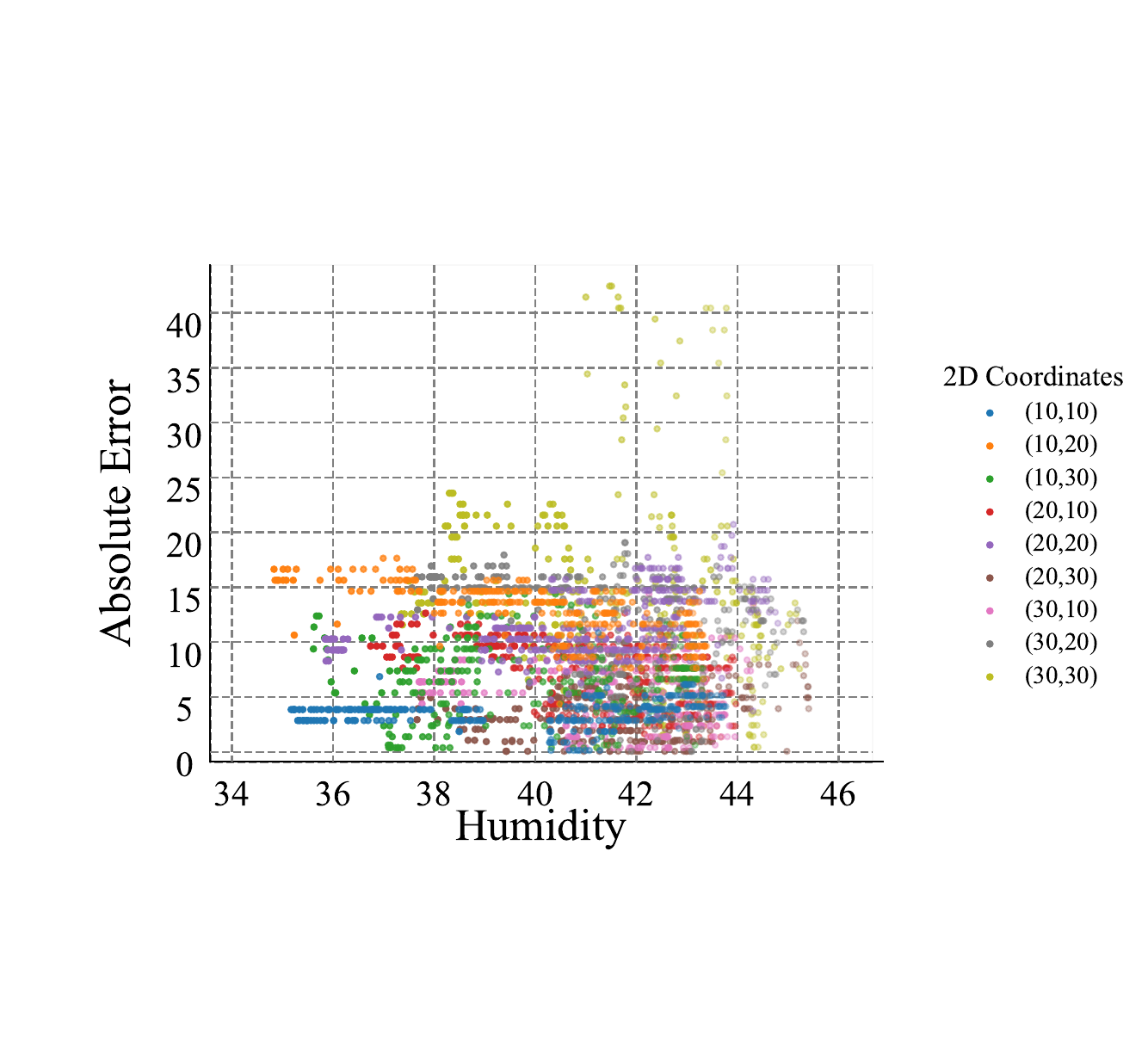}
  \caption{Absolute ranging error vs. humidity.}
  \label{fig:error-vs-humidity}
\end{figure}

To further investigate the impact of the environmental factors on LoRa RF ToF
ranging based on the recorded environmental factors of temperature and humidity,
we present a 3D scatter plot of temperature ($X$)-humidity ($Y$)-absolute
ranging error ($Z$) colored and grouped by RP in \autoref{fig:3d-scatter-plot}
and its projection onto the $X$-$Z$ and $Y$-$Z$ planes in
Figs.~\ref{fig:error-vs-temperature} and \ref{fig:error-vs-humidity},
respectively.

The points of the 3D scatter plot shown in \autoref{fig:3d-scatter-plot}
concentrate along the dimensions of temperature and humidity, instead of being
randomly scattered. Also, when temperature or humidity increases, the absolute
ranging errors at some RPs show a significant increase, and relatively
concentrated ``error bands'' emerge in different ranges as shown in
Figs.~\ref{fig:error-vs-temperature} and \ref{fig:error-vs-humidity}. This
phenomenon indicates that temperature and humidity are correlated with the
ranging errors, which suggests a potential of compensating for the impact of
environmental factors on LoRa RF ToF ranging.


\subsection{Database Validation}
\label{sec:database-validation}
The results of the data analysis presented in Section~\ref{sec:data-analysis}
demonstrate that the ranging errors show a systematic dependence on
environmental factors of temperature and humidity.
To explore the potential of compensating for the impact of environmental
factors on LoRa RF ToF ranging, we construct a simple DNN model whose input
layer consists of two environmental features, temperature and humidity, and
output layer corresponds to the absolute ranging error, thereby enabling an
end-to-end mapping between environmental factors and error bias. The model
architecture is composed of two fully-connected hidden layers and a regression
output layer. During the training, the Adam optimizer and mean squared error
(MSE) loss function are adopted to ensure stable convergence, while early
stopping~\cite{Prechelt2012} and 5-fold
cross-validation~\cite{10.5555/1005332.1044695} are applied to enhance
generalization performance. The hyperparameter values of the DNN model are
summarized in Table~\ref{tab:dnn-hyperparameters}.
\begin{table}[!htb]
  \centering
  \caption{DNN hyperparameter values.}
  \label{tab:dnn-hyperparameters}
  \begin{tabular}{ll}
    \toprule
    \textbf{Parameter}      & \textbf{Value}                    \\
    \midrule
    Input features          & Temperature, Humidity             \\
    Output target           & Absolute ranging error (\si{\m})  \\
    Hidden layers           & Dense(128, ReLU), Dense(32, ReLU) \\
    Output layer            & Dense(1, Linear)                  \\
    Optimizer               & Adam                              \\
    Learning rate           & 0.001                             \\
    Loss function           & MSE                               \\
    Batch size              & 64                                \\
    Max epochs              & 300                               \\
    Early stopping patience & 25                                \\
    \bottomrule
  \end{tabular}
\end{table}

Table~\ref{tab:cv-results} shows the relatively stable results of the
performance evaluation of the DNN model through 5-fold cross-validation, which
provides the validation outcomes for each fold, including the MSE on the
validation set, the denormalized mean absolute error (MAE), the root mean
squared error (RMSE), and the coefficient of determination ($R^{2}$). The
results indicate that under different environmental conditions, temperature and
humidity exert varying impacts on the ranging error. Overall, the DNN model
demonstrates good adaptability to diverse environmental settings. The mean
values of MAE and RMSE are \SI{3.88}{\m} and \SI{5.03}{\m}, respectively,
further confirming that temperature and humidity significantly affect the
ranging error, and that this impact can be efficiently compensated for by the
DNN model to achieve ``error correction.''
\begin{table}[!htb]
  \centering
  \caption{5-fold cross-validation results.}
  \label{tab:cv-results}
  \begin{tabular}{ccccc}
    \toprule
    \textbf{Fold} & \textbf{val\_MSE} & \textbf{MAE (m)}  & \textbf{RMSE (m)} & \textbf{$R^{2}$}  \\
    \midrule
    1             & 0.740044          & 3.689432          & 4.688753          & 0.116254          \\
    2             & 1.116877          & 4.217955          & 5.550302          & 0.079014          \\
    3             & 1.054927          & 4.130346          & 5.410613          & 0.105412          \\
    4             & 0.767498          & 3.654264          & 4.749471          & 0.137807          \\
    5             & 0.771699          & 3.692932          & 4.759478          & 0.140500          \\
    \midrule
    \textbf{Mean} & \textbf{0.890209} & \textbf{3.876986} & \textbf{5.031723} & \textbf{0.115798} \\
    \textbf{Std}  & 0.161346          & 0.244585          & 0.369836          & 0.022609          \\
    \bottomrule
  \end{tabular}
\end{table}

\section{Conclusions}
\label{sec:conclusions}
In this paper, we have investigated the impact of the environmental factors of
temperature and humidity on the performance of LoRa \SI{2.4}{\GHz} RF ToF
ranging in an outdoor environment, which is based on the ranging data collected
together with temperature and humidity on the sports field at the XJTLU south
campus over three weeks.

Through the data analysis based on the 2D and 3D visualization of ranging
errors, we identify not only a rather significant spatial variability of the
mean absolute ranging errors across RPs but also a systematic dependence of
absolute ranging errors on temperature and humidity. The results of the database
validation based on a simple DNN model mapping between environmental factors and
error bias, on the other hand, demonstrate a potential of compensating for the
impact of environmental factors on LoRa RF ToF ranging, which call for advanced
methods of compensating for the effects of environmental factors on LoRa RF ToF
ranging outdoors.

\section*{Acknowledgment}
This work was supported in part by Xi'an Jiaotong-Liverpool University (XJTLU)
Summer Undergraduate Research Fellowships (under Grant SURF-2025-0217).

\balance 


\end{document}